# STATISTICAL MODELLING OF $f_t$ TO PROCESS PARAMETERS IN 30 NM GATE LENGTH FINFETS

B. Lakshmi and R. Srinivasan

Department of Information Technology
SSN College of Engineering, Kalavakkam – 603 110, Chennai, India
laxmi.balu@yahoo.com
srinivasanr@ssn.edu.in

## ABSTRACT

*This paper investigates the effect of process variations on unity gain frequency ($f_t$) in 30 nm gate length FinFET by performing extensive TCAD simulations. Six different geometrical parameters, channel doping, source/drain doping and gate electrode work function are studied for their sensitivity on $f_t$. It is found that $f_t$ is more sensitive to gate length, underlap, gate-oxide thickness, channel and Source/Drain doping and less sensitive to source/drain width and length, and work function variations. Statistical modelling has been performed for $f_t$ through design of experiment with respect to sensitive parameters. The model has been validated through a comparison between random set of experimental data simulations and predicted values obtained from the model.*

## 1. INTRODUCTION

The progress in CMOS technology has made it well suited for RF and microwave operations at high level of integration [1], and the continuous improvement of the device performance has made it a contender for low-power and, eventually, low-cost radio front end. In the area of multi-gate transistors, double-gate FinFETs are considered a serious contender for channel scaling [2], [3] because of their quasi-planar structure and the compatibility with CMOS. Several authors have already studied the low field mobility in fins of various widths [4]-[6]. RF performance of FinFETs is reported in [7], [8].

Unity gain frequency ($f_t$) is one of the important metric in RF applications. $f_t$ is defined as the frequency at which the current gain of the device becomes unity. $f_t$ is calculated by

$$f_t = \frac{g_m}{2\pi C_{gg}} \quad (1)$$

where $C_{gg}$ is the combination of $C_{gs}$, $C_{gd}$, overlap capacitance and any other fringing capacitance. In this article, nine different geometrical parameters related to FinFET are varied to capture their sensitivity on $f_t$.

This paper analyzes double-gate FinFETs as a downscaling option for CMOS technology from an RF perspective. The effect of various structural and doping parameters (9 parameters in total) on $f_t$ is studied and the five most sensitive parameters are identified. Using these sensitive parameters a 5 point DOE (design of experiments) is designed and the simulations are done. i.e. this work is based on design and simulation of the nominal device, design of experiment and running of the experiment, extraction of results, fitting the response surfaces (models) and testing of the models. We have modelled $f_t$ in terms of the most sensitive parameters like gate length, underlap, gate oxide thickness, channel and source/drain doping. The model that describe these quantities have an approximated form [9] as,

$$y = b_0 + b_1x_1 + b_2x_2 + b_3x_3 + b_4x_4 + b_5x_5 + b_{11}x_1^2 + b_{22}x_2^2 + b_{33}x_3^2 + b_{44}x_4^2 + b_{55}x_5^2 \\ + b_{12}x_1x_2 + b_{13}x_1x_3 + b_{14}x_1x_4 + b_{15}x_1x_5 + b_{23}x_2x_3 + b_{24}x_2x_4 + b_{25}x_2x_5 \\ + b_{34}x_3x_4 + b_{35}x_3x_5 + b_{45}x_4x_5 \quad (2)$$

where $x_1$ is the gate length, $x_2$ is the underlap, $x_3$ is the gate oxide thickness, $x_4$ is the channel doping and $x_5$ is the Source/Drain doping, y is the unity gain frequency, b's are the fitting parameters determined by the data obtained from the experiment. Next section deals with the simulation methodology followed in this paper. Section III discusses the simulation results and statistical modelling. Finally section IV provides conclusions.

## 2. SIMULATION METHODOOGY

Sentaurus TCAD simulator from Synopsys [10] is used to perform all the simulations. This simulator has many modules and the following are used in this study.

- Sentaurus structure editor (SDE): To create the device structure, to define doping, to define contacts, and to generate mesh for device simulation
- Sentaurus device simulator (SDEVICE): To perform all DC and AC simulations
- Inspect and Tecplot: To view the results.

The physics section of SDEVICE includes the appropriate models for band to band tunnelling, quantization of inversion layer charge, doping dependency of mobility, effect of high and normal electric fields on mobility, and velocity saturation. The structure generated from SDE is shown in Fig 1. Doping and mesh information can also be observed in Fig. 1. Figure 2 shows the schematic diagram of the device. Totally nine different parameters are considered in this study. Out of them, six are geometrical parameters - gate length ($L_g$), underlap ($L_{un}$), fin width (W), source/drain width (SW), source/drain length (SL), and $T_{ox}$ and these are shown in Fig. 2. Other three parameters are channel doping ($N_{ch}$), source/drain doping ($N_{SD}$) and gate electrode work function (WF). The various process parameters considered in this study and their range are given in Table. 1. Table 1 also gives the dimensions of the nominal device. Standard AC simulations are done in SDEVICE and $f_t$ is extracted from these results. $f_t$ is the frequency at which |Y21/Y11| equals one, and it strongly depends on the gate bias. At various gate biases $f_t$ is calculated and the maximum of them is taken as $f_t$. Supply voltage ($V_{dd}$) used in this study is 0.8 V.

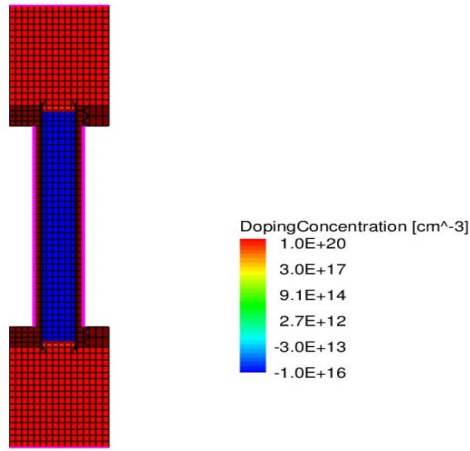

Figure 1. Structure of the Dual-Gate FinFET

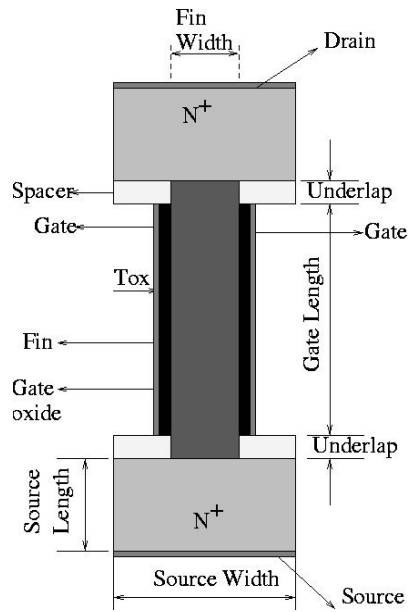

Figure 2. Schematic view of Dual-Gate FinFET

Table 1. Characteristics of the device

| Process parameter | Nominal value | Range |
|---|---|---|
| Gate length ($L_g$) | 30 nm | 20 nm to 40 nm |
| Underlap ($L_{un}$) | 3 nm | 1 nm to 8 nm |
| Fin Width (W) | 4 nm | 2 nm to 7 nm |
| Source Length (SL) | 15 nm | 10 nm to 20 nm |
| Source Width (SW) | 8 nm | 4 nm to 16 nm |
| Channel Doping ($N_{ch}$) | $1\times10^{16}/cm^3$ | $1\times10^{15}/cm^3$ to $1\times10^{19}/cm^3$ |
| Source Drain Doping ($N_{SD}$) | $1\times10^{20}/cm^3$ | $1\times10^{18}/cm^3$ to $2x^{20}/cm^3$ |
| Oxide Thickness ($T_{ox}$) | 1 nm | 0.5 nm to 2 nm |
| Gate Work Function (WF) | 4.337 eV | 4.13 eV to 4.9 eV |

## 3. RESULTS AND DISCUSSION

### 3.1 Sensitivity Analysis of Process Parameters

The nine different process parameters are varied one at a time, according to the range given in Table 1 and their sensitivity to $f_t$ is analysed in this section.

### 3.1.1 Variation in Gate Length

Figure 3 shows the variation of $f_t$ against $L_g$. It can be observed from Fig. 3 that $f_t$ initially increases and then decreases. As per (1), $f_t$ is decided by both $g_m$ and $C_{gg}$. While $g_m$ degrades with $L_g$, $C_{gg}$ shows a different behaviour i.e. initially decreases with $L_g$ and then increases (Fig 4). The initial decrease of $C_{gg}$ can be attributed to the reduction of $C_{gd}$ [11]. At some point, $C_{gs}$ starts dominating and calls for the increase in $C_{gg}$. The combined behaviour of $g_m$ and $C_{gg}$ contributes to the variation of $f_t$ w.r.t. $L_g$.

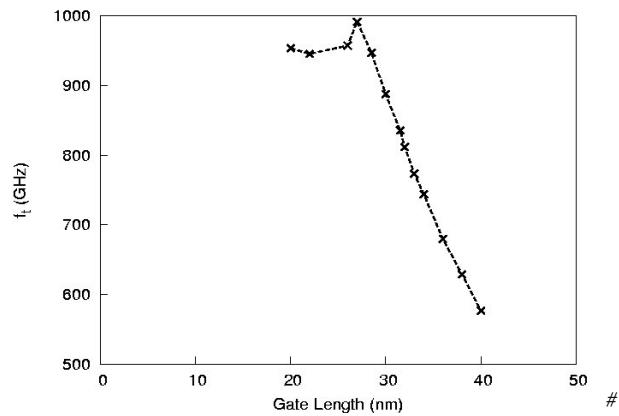

Figure 3. Variation of $f_t$ with respect to $L_g$

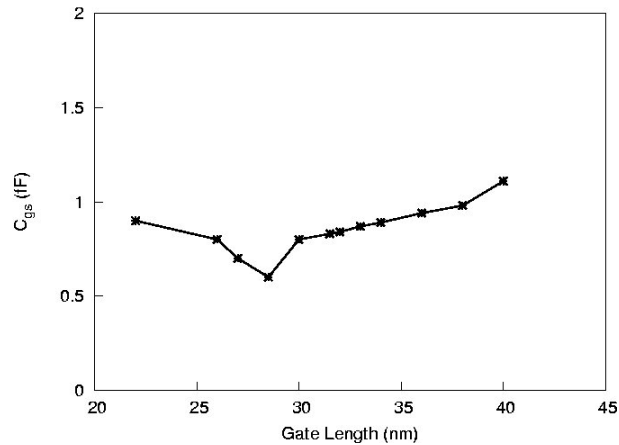

Figure 4. Variation of $C_{gs}$ with respect to $L_g$

### 3.1.2. Variation in Underlap

Figure 5 depicts the plot between $f_t$ and $L_{un}$. It can be seen from Fig 5 that $f_t$ initially increases and then decreases w.r.t. $L_{un}$. Increasing $L_{un}$ reduces the fringing capacitance, and thereby decreases $C_{gg}$ [12], [13]. The $C_{gg}$ in DGMOS can be expressed as

$$C_{gg} = \text{Series}\left(C_{ox}, C_{si}\right) \| C_{ov} \| C_{fringing} \qquad (3)$$

where $C_{ox}$ is the oxide capacitance, $C_{si}$ is the silicon body capacitance, $C_{ov}$ is the gate to source/drain overlap capacitance and $C_{fringing}$ is the fringing capacitance and is given by

$$C_{fringing} = \frac{WK\epsilon_{di}}{\pi} \ln \frac{\pi W}{\sqrt{L_{un}^2 + T_{ox}^2}} e^{-\left|\frac{L_{un} - T_{ox}}{L_{un} + T_{ox}}\right|} \qquad (4)$$

When $L_{un}$ increases current degrades and thereby $g_m$ monotonically decreases. The combined behavior of $g_m$ and $C_{gg}$ is responsible for the $f_t$ trend seen in Fig. 5.

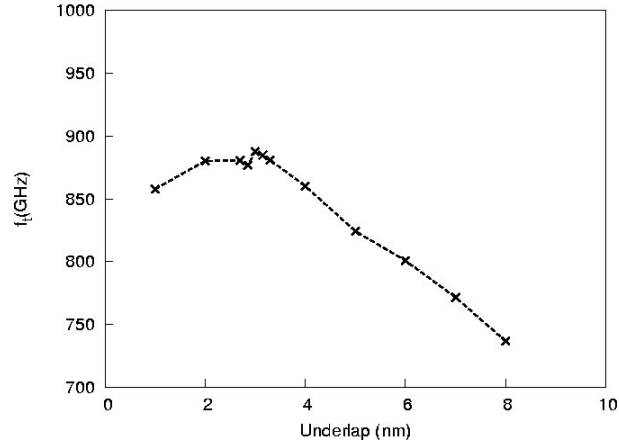

Figure 5. Variation of $f_t$ with respect to $L_{un}$

### 3.1.3. Variation in Fin Width

When W is varied we may either face volume inversion or may not, depending upon the channel doping levels. When the channel doping is $1 \times 10^{16}/cm^3$ volume inversion is not seen [14]. Therefore, the increase in W increases current and thereby $g_m$ and $f_t$. Figure 6 shows this kind of behaviour between $f_t$ and W. For the channel doping around $1.5 \times 10^{18}/cm^3$, volume inversion effect is seen which causes $f_t$ to decrease initially and then to increase w.r.t W. This is depicted in Fig. 7.

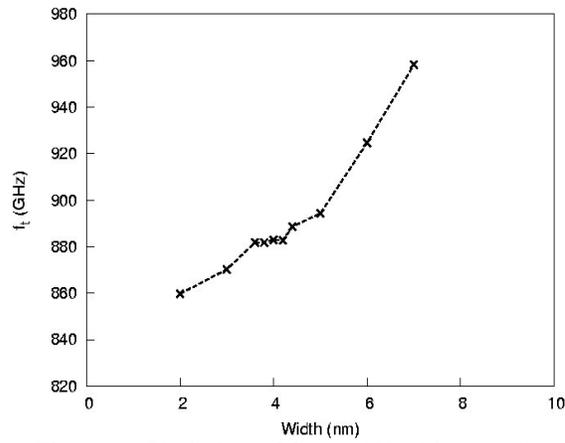

Figure 6. Variation of $f_t$ w.r.t W without volume inversion

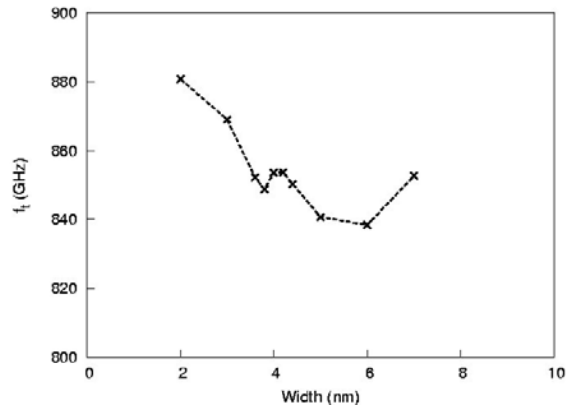

Figure 7. Variation of $f_t$ w.r.t W with volume inversion

### 3.1.4. Variation in Source Length

Figure 8 shows the $f_t$ versus source length plot. Increasing source length increases the parasitic resistance associated with the channel and degrades $g_m$ which in turn decreases $f_t$.

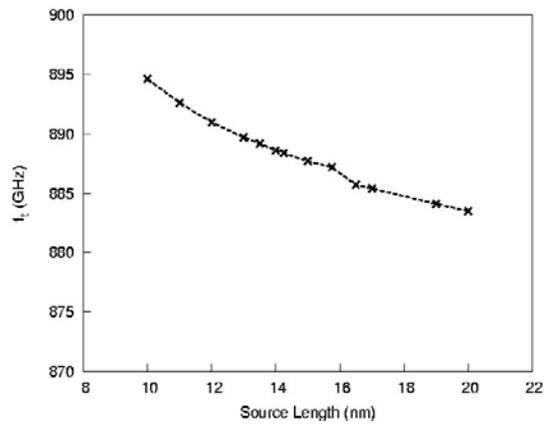

Figure 8. Variation of $f_t$ with respect to SL

### 3.1.5. Variation in Source Width

Figure 9 shows $f_t$ as a function of source width. It can be noticed from Fig. 9 that $f_t$ is almost independent of SW.

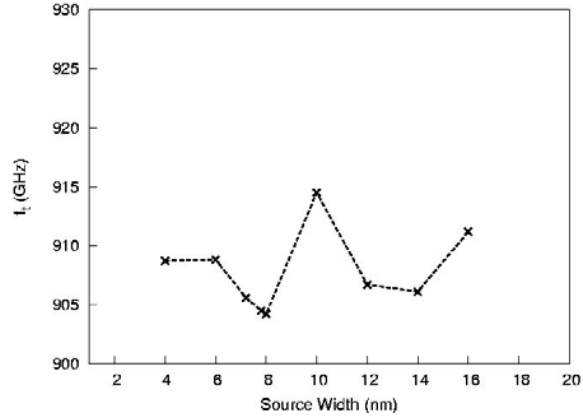

Figure 9. Variation of $f_t$ with respect to SW

### 3.1.6. Variation in Oxide Thickness

Figure 10 shows the variation of $f_t$ with $T_{ox}$. $f_t$ increases initially w.r..t $T_{ox}$. Both $g_m$ and $C_{gg}$ together control $f_t$. $C_{gg}$ always decreases when $T_{ox}$ increases whereas $g_m$ may go down or high depending on whether we are driven by short channel effects or not. The combined effect of $g_m$ and $C_{gg}$ decides $f_t$ behaviour with respect to $T_{ox}$.

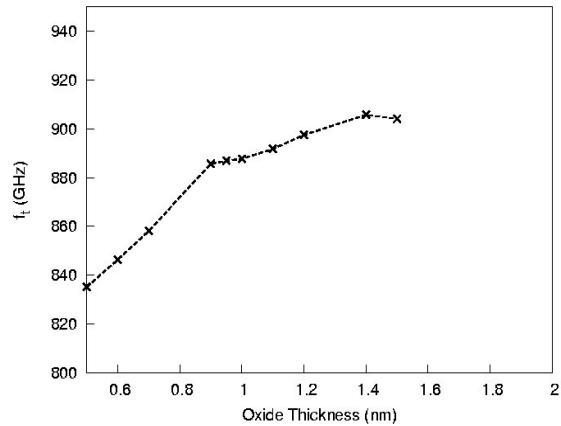

Figure 10. Variation of $f_t$ with respect to $T_{ox}$

### 3.1.7. Variation in Channel Doping

Figure 11 shows the variation of $f_t$ against $N_{ch}$. Threshold voltage of DGFET/FinFET is insensitive up to $1 \times 10^{17}/cm^3$ [15]. From which it can be reasoned out that $f_t$ is also insensitive at lower channel doping levels. The same is seen in Fig. 11. At higher doping levels, $f_t$ decreases due to $g_m$ degradation.

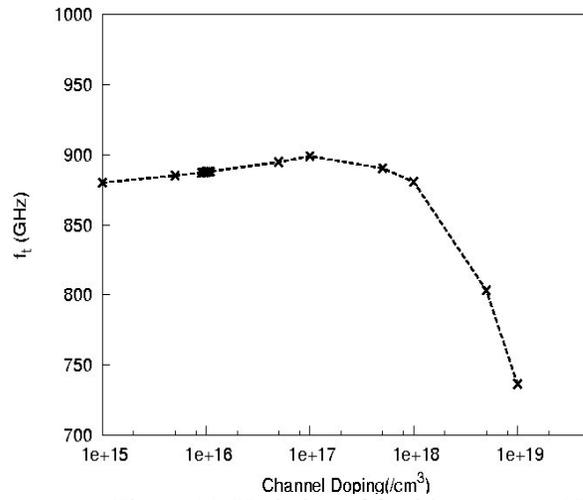

Figure 11. Variation of $f_t$ with respect to $N_{ch}$.

### 3.1.8. Variation in Source/drain Doping

Figure 12 shows the variation of $f_t$ with source/drain doping. When $N_{SD}$ increases $I_{on}$ and $g_m$ increase due to the lowered parasitic series resistance values, and thereby $f_t$ increases. This is reflected in Fig.12

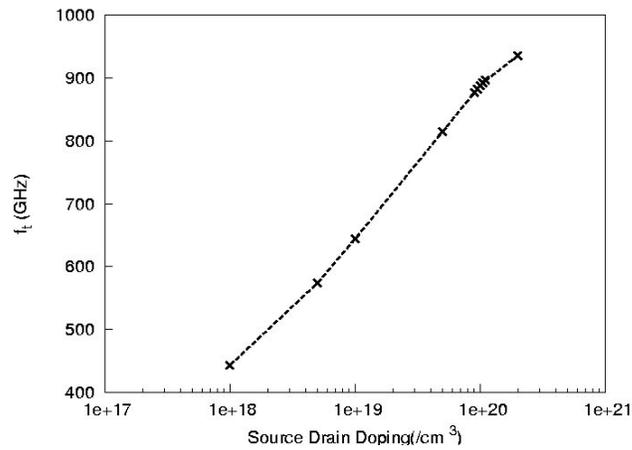

Figure 12. Variation of $f_t$ with respect to $N_{SD}$

### 3.1.9. Variation in Work Function

High frequency characteristics are less sensitive to work function variation [16]. Therefore, $f_t$ is expected to be indifferent to gate electrode work function. Figure 13 shows $f_t$ versus gate work function plot and it can be noticed that $f_t$ exhibits a flat behaviour w.r.t work function.

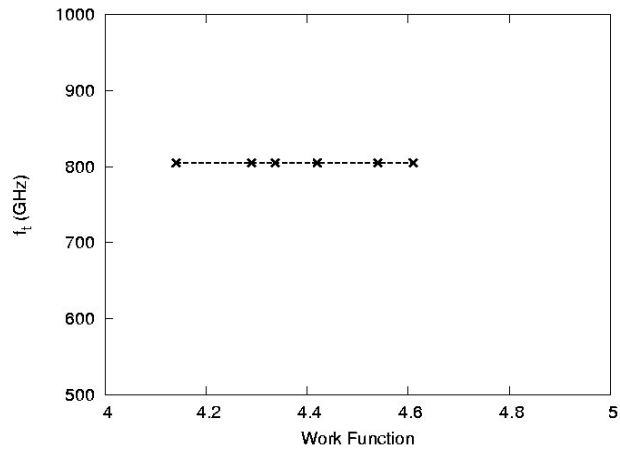

Figure 13. Variation of $f_t$ with respect to WF

## 3.2 Statistical modelling:

The sensitive parameters are chosen as $L_g$, $L_{un}$, $T_{ox}$, $N_{ch}$ and $N_{SD}$. These sensitive parameters are made to undergo a variability study with the help of design of experiments. The range of these sensitive parameters for the variability study is given in Table 2 which results in 2500 simulations. The model that has been obtained by the regression technique is generated with the help of SPSS [17].The regression coefficients for the second order polynomial of the process parameters are shown in Table 3.

Table 2 .Range of Sensitive Parameters

| Process parameters | Range | Points in the range |
| --- | --- | --- |
| Gate length ($L_g$) | 20 nm to 40 nm | 20, 25, 30, 35 and 40 (nm) |
| Underlap ($L_{un}$) | 1nm to 9 nm | 1, 3, 5, 7 and 9 (nm) |
| Oxide Thickness ($T_{ox}$) | 0.5 nm to 2.2 nm | 0.5, 1, 1.2, 2 and 2.2 (nm) |
| Channel Doping ($N_{ch}$) | $1 \times 10^{16}$ /cm$^3$ to $1 \times 10^{19}$ /cm$^3$ | $1 \times 10^{16}$, $1 \times 10^{17}$, $1 \times 10^{18}$ and $1 \times 10^{19}$ (/cm$^3$) |
| Source Drain Doping ($N_{SD}$) | $5 \times 10^{18}$/cm$^3$ to $2 \times 10^{20}$ /cm$^3$ | $5 \times 10^{18}$, $5 \times 10^{19}$, $9.5 \times 10^{19}$, $1.1 \times 10^{20}$ and $2 \times 10^{20}$ (/cm$^3$) |

\

Table 3. Process parameters along with their corresponding regression coefficients

| Factor | Coefficient | Factor values |
|---|---|---|
| Constant | $b_o$ | 2.765E12 |
| $L_g$ | $b_1$ | -1.103E20 |
| $L_{un}$ | $b_2$ | -3.964E19 |
| $T_{ox}$ | $b_3$ | 4.007E19 |
| $N_{ch}$ | $b_4$ | -.052 |
| $N_{SD}$ | $b_5$ | .009 |
| $L_g^2$ | $b_{11}$ | 1.164E27 |
| $L_{un}^2$ | $b_{22}$ | -4.210E27 |
| $T_{ox}^2$ | $b_{33}$ | -3.564E28 |
| $N_{ch}^2$ | $b_{44}$ | 9.698E-16 |
| $N_{SD}^2$ | $b_{55}$ | -1.489E-17 |
| $L_g L_{un}$ | $b_{12}$ | 1.874E27 |
| $L_g T_{ox}$ | $b_{13}$ | 1.639E27 |
| $L_g N_{ch}$ | $b_{14}$ | 1266894.540 |
| $L_g N_{SD}$ | $b_{15}$ | -100967.539 |
| $L_{un} T_{ox}$ | $b_{23}$ | 7.137E27 |
| $L_{un} N_{ch}$ | $b_{24}$ | -3105744.884 |
| $L_{un} N_{SD}$ | $b_{25}$ | -124191.069 |
| $T_{ox} N_{ch}$ | $b_{34}$ | -1672454.265 |
| $T_{ox} N_{SD}$ | $b_{35}$ | 23534.568 |
| $N_{ch} N_{SD}$ | $b_{45}$ | -4.254E-17 |

In order to study the statistical nature of the device output with respect to sensitive parameters, we have generated $3^5$ (=243) uniformly distributed pseudo-random numbers for each of these sensitive parameters and $f_t$ values are predicted from the generated model. For the same set of random numbers generated, TCAD simulations are carried out and the $f_t$ values are extracted. To have a correlation plot TCAD values are plotted against model values and the same is depicted in Fig. 14. The correlation coefficient is found out as r=0.992.

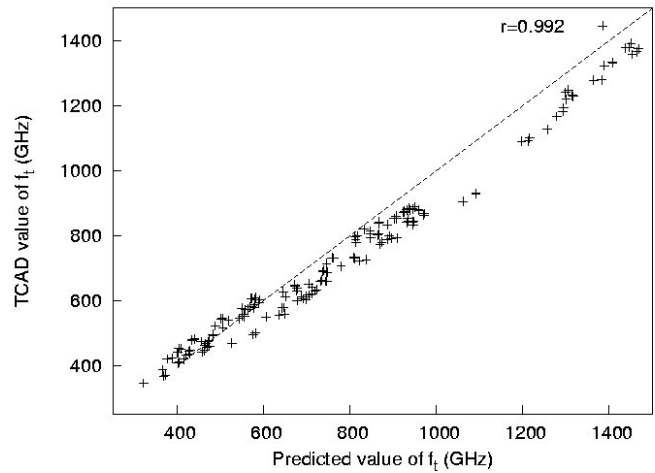

Fig. 14: Correlation plot between TCAD simulated data and model predicted data for $f_t$. Correlation coefficient r is depicted here

## 4. CONCLUSION

Six geometrical parameters ($L_g$, $L_{un}$, W, SW, SL, and $T_{ox}$), and three non-geometrical parameters ($N_{ch}$, $N_{SD}$, and WF) have been varied over a range and their effect on $f_t$ have been studied through TCAD simulations. It was found that $L_g$, $L_{un}$, $T_{ox}$, $N_{ch}$ and $N_{SD}$ are more sensitive parameters whereas gate electrode work function, source/drain length and width are less sensitive parameters. By running DOE using the sensitive parameters in TCAD, statistical modelling has been performed. Correlation coefficient around 0.992 was got while running the simulations with random set of values for sensitive parameters.

## ACKNOWLEDGEMENT

This work is supported by Department of Science & Technology, Government of India under SERC scheme

## REFERENCES

[1]  L. Larson, "Silicon technology tradeoffs for radio-frequency/mixed signal systems-on-a-chip," IEEE Trans. Electron Devices, vol. 50, no. 3, pp. 683–699, Mar. 2003.

[2]  International Technology Roadmap for Semiconductor (ITRS). [Online]. Available: www.itrs.net/Links/2005ITRS/PIDS2005.pdf

[3]  D. J. Frank, R. H. Dennard, E. Nowak, P. M. Solomon, Y. Taur, and H. S. P. Wong, "Device scaling limits of Si MOSFETs and their application dependencies," Proc. IEEE, vol. 89, no. 3, pp. 259-288, Mar. 2001.

[4]  F. Gamiz, J. Roldan, J. Lopez-Villanueva, P. Cartujo-Cassinello, and J. E. Carceller, "Surface roughness at the Si?SiO2 interfaces in fully depleted silicon-on-insulator inversion layers," J. Appl. Phys., vol. 86, no. 12, pp. 6854-6863, Dec. 1999.

[5]     A. Khakifirooz and D. Antoniadis, "On the electron mobility in ultrathin SOI and GOI," IEEE Electron Device Lett., vol. 25, no. 2, pp. 80-82, Feb. 2004.

[6]     K. Uchida, H. Watanabe, J. Koga, A. Kinoshita, and S. Takagi, "Experimental study on carrier transport mechanism in ultrathinbody SOI MOSFETs," in Proc. IEEE SISPAD, Sep. 2003, pp. 8–13.

[7]     S. Nuttinck, "Ultra-thin-body silicon-on-insulator as a CMOS downscaling option: An RF perspective," IEEE Trans. Electron Devices,vol. 53, no.5, pp.   1193–1199, May 2006.

[8]     D. Lederer, B. Parvais, A. Mercha, N. Collaert, M. Jurczak, J.-P. Raskin, and S. Decoutere, "Dependence of finFET RF performance on fin width," in Proc. 6th Top. Meeting SiRF, San Diego, CA, Jan. 18–20, 2006, pp. 4–6.

[9]     Montgomery DC. Design and analysis of experiments. 5th ed. New York: John Wiley & Sons; 2001.

[10]    Synopsys Sentaurus Device User Guide, 2008-09

[11]    Han-Su Kim, Kangwook Park, Hansu Oh, and Eun Seung Jung" Importance of Vth and Substrate Resistance Control for RF Performance Improvement in MOSFETs" IEEE Electron Device letters, Vol. 30, No. 10, October 2009

[12]    Fathipour Morteza, Nematian Hamed, Kohani Fatemeh " The Impact of Structural Parameters on the electrical characteristics of Nano scale DC-SOI MOSFETs in sub-thereshold region" SETIT 2007 4th International Conference: Sciences of Electronic, Technologies of Information and Telecommunications March 25-29, 2007

[13]    R. Shrivastava and K. Fitzpartick, "A simple model for the overlap capacitance of a VLSI MOS device." IEEE Trans.Electron Devices, Vol. ED-29, pp.1870-1875,1982

[14]    G Curatola, S. Nuttinck, "The Role of Volume Inversion on the Intrinsic RF Performance of Double-Gate FinFETs", IEEE Transactions On Electron Devices, Vol. 54, No.1, Jan 2007.

[15]    Shiying Xiong and Jeffrey Bokor," Sensitivity of Double-Gate and FinFET Devices to Process Variations" IEEE Transactions on Electron Devices, Vol. 50, No. 11, Nov 2000.

[16]    Chih-Hong Hwang, Tien-Yeh Li1, Ming-Hung Han, Kuo-Fu Lee, Hui-Wen Cheng1, and Yiming Li," Statistical Analysis of Metal Gate Workfunction Variability, Process Variation, and Random Dopant Fluctuation in Nano-CMOS Circuits", Proceedings of 14th ¬International Conference on Simulation of Semiconductor Processes and Devices," (SISPAD), Pg No 99-102, 2009.

[17]    SPSS Base 15.0 User's Guide, 2006.